\documentclass[%
 reprint,
 amsmath,amssymb,
 aps,
 pre
]{revtex4-1}

\usepackage{graphicx}
\usepackage{dcolumn}
\usepackage{bm}
\usepackage{float}
\usepackage{listings}
\usepackage{multirow}
\usepackage{color}

\begin{document}


\title{Clique Densification in Networks}
\author{Haochen Pi$^{1}$}
\author{Keith Burghardt$^{2}$}
\author{Allon G.\ Percus$^{3,2}$}
\author{Kristina Lerman$^{2}$}
\affiliation{$^{1}$ 
Department of Computer Science,
University of Southern California, Los Angeles, CA 90007, USA
}
\affiliation{$^{2}$
Information Sciences Institute, University of Southern California, Marina del Rey, CA 90292, USA
}
\affiliation{$^{3}$
Institute of Mathematical Sciences, Claremont Graduate University, Claremont, CA 91711, USA
}

\date{\today}

\begin{abstract}
Real-world networks are rarely static. Recently, there has been increasing interest in both network growth and network densification, in which the number of edges scales superlinearly with the number of nodes. Less studied but equally important, however, are scaling laws of higher-order cliques, which can drive clustering and network redundancy. In this paper, we study how cliques grow with network size, by analyzing several empirical networks from emails to Wikipedia interactions. Our results show superlinear scaling laws whose exponents increase with clique size, in contrast to predictions from a previous model. We then show that these results are in qualitative agreement with a new model that we propose, the Local Preferential Attachment Model, where an incoming node links not only to a target node but also to its higher-degree neighbors. Our results provide new insights into how networks grow and where network redundancy occurs.
\end{abstract}

\pacs{Valid PACS appear here}
\maketitle


\section{Introduction}

Networks underlie a wide variety of social phenomena, from the spread of disease and information \cite{Holme2012} to the formation of collaborations \cite{Burghardt2021,Leskovec2007}. The evolution of networks has been a popular research topic since the Barabasi-Albert model demonstrated that growth through preferential attachment can explain a fundamental property of networks---their heavy-tailed degree distributions \cite{BA,Jeong2003}. More recent research has studied another fundamental aspect of network growth, known as densification, where the number of links increases super-linearly with the number of nodes \cite{Leskovec2007}.  Densification can create advantages for larger systems: for instance, in collaboration networks, it provides more opportunities for researchers at larger institutions over smaller ones \cite{Burghardt2021}. Several network growth models have been developed to help explain mechanisms of specific networks, such as gene regulatory networks \cite{Teichmann2004,Foster2006}, or provide general mechanisms of patterns seen in empirical data, such as fitness \cite{Bell2017}, graph spectra \cite{Kunegis2010}, or copying mechanisms \cite{Bhat2016,Lambiotte2016,Burghardt2021}, among others \cite{Dudin2014,zalanyi_properties_2003,paranjape_motifs_2017,leskovec_microscopic_2008}.

Growth of higher-order structures in networks is a less studied aspect of network growth. Although some higher-order structures, such as triangles \cite{Watts1998}, have long been known to play an important role in network phenomena, less attention has been devoted to how these and higher-order motifs form in growing networks. Recent research, notably by Bhat et al.~\cite{Bhat2016} and Lambiotte et al.~\cite{Lambiotte2016}, has offered potential mechanisms that predict how edges and 
larger cliques will scale as a function of network size. (For clarity, a clique of size $k$ is a fully connected subgraph, with $k$ nodes and $k(k - 1)/2$ edges). The mechanism of clique formation proposed by Lambiotte et al., however, has not been tested empirically before.

\begin{figure*}[tbh!]
    \includegraphics[width=\textwidth]{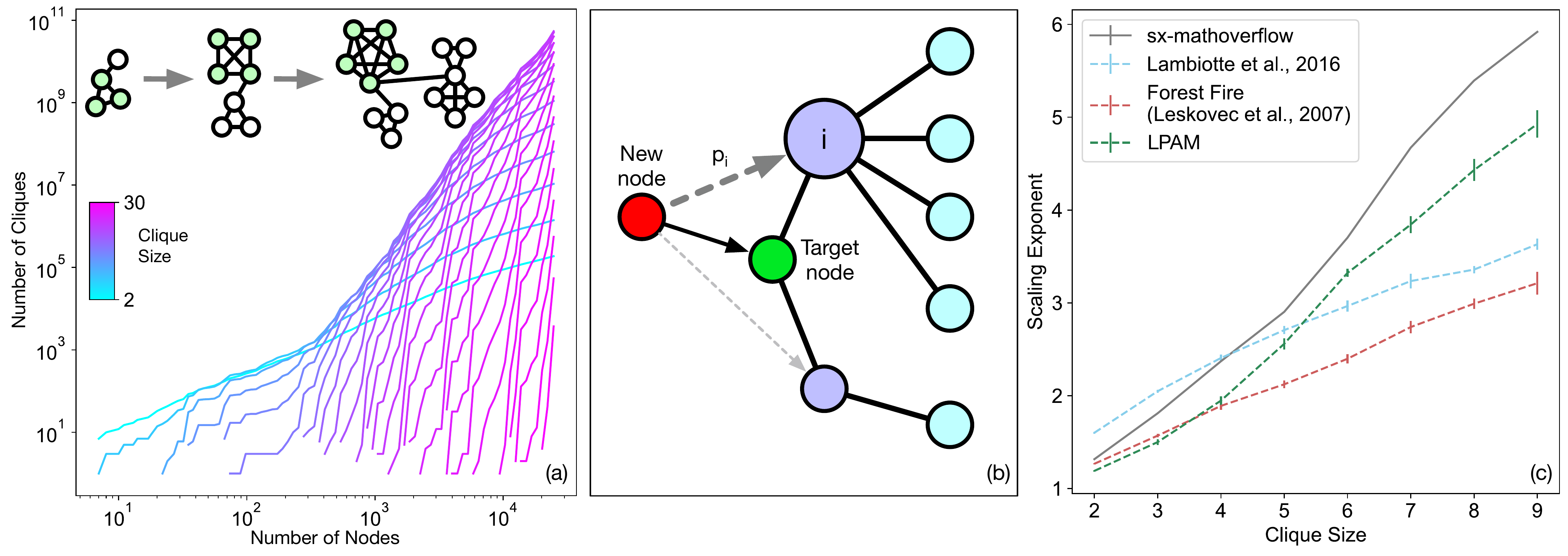}
     \caption{(a) Number of cliques of a given size vs. the number of nodes in the network for the Math Overflow question-answer website~\cite{paranjape_motifs_2017}. See also Supplementary Figure~S1 for results on other networks.
     (b) Local Preferential Attachment Model (LPAM) preferentially attaches to higher-degree neighbors of the target node. A new node (red) connects to a random target node (green), as well as to the target node's higher-degree existing neighbors (purple).
     (c) Scaling laws versus clique size for LPAM, node copying mechanism of Lambiotte at al \cite{Lambiotte2016}, Forest Fire model \cite{Leskovec2007}, and Math Overflow data.
     }\label{fig:fig1}
 \end{figure*}

In this paper, we study clique formation in growing networks. 
Fig.~{\ref{fig:fig1}}a considers the case of an empirical network of user interactions on the question-answer website known as Math Overflow~\cite{paranjape_motifs_2017}  (answers to questions, comments to questions, and comments to answers). Plotting the number of cliques of differing size $k$ as a function of network size (measured by the number of nodes), we see that the number of edges ($k=2$) grows super-linearly with network size. Network degree therefore increases with network size, consistent with previous results~\cite{Lambiotte2016,Bhat2016,Burghardt2021,Leskovec2007}. But crucially, we observe that the number of triangles ($k=3$) and larger cliques grows \emph{even faster}, leading to an increased level of redundant connections in the network. This effect, which we call \emph{clique densification}, is found in many empirical networks (see also Supplementary Figure~S1). We also find that these networks form links locally, i.e.,\ between nearby nodes, and preferentially connect to high-degree nodes. Furthermore, the effect of 2-cliques being overtaken by increasingly large clique sizes in Fig.~{\ref{fig:fig1}}a gives rise to an intriguing envelope structure that itself appears to follow a power law.

In order to explain our findings, we propose the Local Preferential Attachment Model (LPAM) that combines two prevalent mechanisms in networks: copying (linking not only to a target but also to some of its neighbors)~\cite{Krapivsky2005}, and preferential attachment (linking preferentially to higher-degree nodes). While copying alone can explain some network densification, it does not explain why the representation of large cliques grows so rapidly in networks. Similarly, preferential attachment cannot explain densification at all. The two mechanisms together, however, are key to understanding how such dense substructures arise in networks. These substructures can be useful, for example, when links are removed because they provide redundancy that maintains network connectivity. This may help us understand seeming inefficiencies in network formation, as the density of these subgraphs may preserve the giant connected component of a network. Moreover, the copying and preferential attachment mechanisms could assist in explaining the formation of dense subcommunities in networks~\cite{Chen2012}. Our work provides a new understanding of how network structure evolves and can help account for these behaviors.

\section{Methods}\label{Methods}
In this section, we describe how cliques in a network are counted.  We define our model that better explains how cliques scale super-linearly with network size, and we discuss how we fit this and other models to data.

\subsection{Empirical Networks}
The empirical datasets we use are freely available from the SNAP library \cite{snapnets}. We take 11 graphs that contain temporal information, ignoring weights and edge direction: College Messages \cite{Panzarasa2009} (nodes are users, and edges are messages between individuals), an email network at a large European institution \cite{paranjape_motifs_2017} (nodes are users and edges are emails between users), Reddit hyperlinks within the body and within the title of posts \cite{kumar2018community} (nodes are users and edges are 
links to comments between users), Bitcoin Alpha and Bitcoin OTC trust weighted signed networks \cite{kumar2016edge,kumar2018rev2} (nodes are users and edges represent degree of trust, where we ignore the edge sign), and conversations on Ask Ubuntu, Math Overflow, Stack Overflow, Stack Exchange Super User boards (nodes are users and edges represent comments to questions or answers, or answers to questions between users) \cite{paranjape_motifs_2017}, and Wikipedia's talk pages (nodes are users and edges are comments between users) \cite{paranjape_motifs_2017}. Data are captured cumulatively, such that links and nodes will appear but not disappear from the first to the last timestamp.

\subsection{Counting Cliques}
It is typically a challenge to analyze high-order network properties, such as cliques, in part because finding the largest clique in a network is NP-hard \cite{Rossi2014}. Pivoter \cite{jain_power_2020}, however, helps speed up clique counting, allowing clique densification to be studied. Pivoter is based on the Succinct Clique Tree, which efficiently stores a representation of all cliques in the network. This is built via an algorithm called pivoting, which reduces the recursion tree used to find the cliques. We use this method to study all empirical networks. Code used to model and analyze data is available at \url{https://github.com/haochenpi314/Clique-Densification}.

\subsection{Local Preferential Attachment Model}

We find three attributes of growing networks that we aim to capture within a single mechanistic model: (a) the number of cliques scaling super-linearly with the network size, (b) nodes forming new links with nearby nodes, and (c) nodes preferentially connecting to high-degree nodes. One theoretically grounded mechanistic model is by Lambiotte et al. \cite{Lambiotte2016}, in which nodes enter the network, find a random target node to connect with, and then also connect to random neighbors of that target node. Their model provides theoretical predictions on the scaling laws of edges and higher-order cliques versus network size, but does not assume any preferential attachment mechanism. 

We therefore expand on this model with LPAM, shown in Fig.~{\ref{fig:fig1}}b. Consider a process where, at each time step, a new node (red node) enters the network.  It connects to an existing target node (green node) 
chosen uniformly at random, and also connects to some number of neighbors (purple nodes) of the target, with preference given to higher-degree nodes (larger-sized nodes in Fig.~{\ref{fig:fig1}}b).  
%

LPAM is characterized by two parameters, $p$ and $r$. For a target node of degree $k$, the marginal probability of establishing a connection to a given one of its neighbors is $p$, such that the expected number of new connections is $pk$. However, conditional on the neighbor's own degree, this probability depends on $r$.  The parameter $r$ interpolates linearly between the case of no preferential attachment at all ($r=0$), corresponding to the Lambiotte et al. model~\cite{Lambiotte2016}, and the case of strong preferential attachment ($r=1$).  Specifically, for the $i$th neighbor of the target node, we define the initial scaled probability
\begin{equation}
    p_i=p\,\frac{k_i}{\sum_{j=1}^{k} k_j/k},
\end{equation}
where $k_i$ is the degree of the $i$th neighbor.  If $p_i$ exceeds a threshold level $p+(1-p)r$, then the ``excess'' probability $p_i - (p+(1-p)r)$ is spread over the probabilities of connecting to other neighbors of the target node, giving new probabilities $p_j'=p_j + (p_i - (p+(1-p)r))/(k-1)$. As this may result in certain probabilities exceeding $p+(1-p)r$, the process is iterated until all probabilities fall below that threshold. The end result is an expectation value independent of $r$: we continue to connect to $pk$ nodes on average, but with a preferential attachment to higher-degree nodes. The threshold level allows us to smoothly transition between strictly preferential attachment ($r=1$) and the Lambiotte et al.\ node copying model \cite{Lambiotte2016}.

For a network with $N$ nodes and $L(N)$ links, the network growth mechanism implies, as in~\cite{Lambiotte2016},
\begin{equation}
    L(N+1) = L(N) + 1 + 2p\,\frac{L(N)}{N}.
\end{equation}
Following the same theory as in Bhat et al.~\cite{Bhat2016}, this results in
\begin{equation}
L(N)=\begin{cases}
N/(1-2p) & p<1/2\\
N \ln N & p = 1/2\\
A(p) N^{2p} & p > 1/2
\end{cases}  
\end{equation}
i.e.,\ the number of links scales superlinearly for $p>1/2$, where $A(p) = [(2p - 1)\Gamma(1 + 2p)]^{-1}$.
Sadly, because $p_{i}$ depends on the other neighbor degrees $k_j$, higher-order dependencies are not solvable, such as the number of triangles as a function of $N$. We instead calculate scaling laws numerically by taking a linear fit of the log of the number of cliques versus log of the network size (see Fig.~{\ref{fig:fig1}}a) for different realizations of this model. 


\begin{figure*}[bth]
    \includegraphics[width=0.9\linewidth]{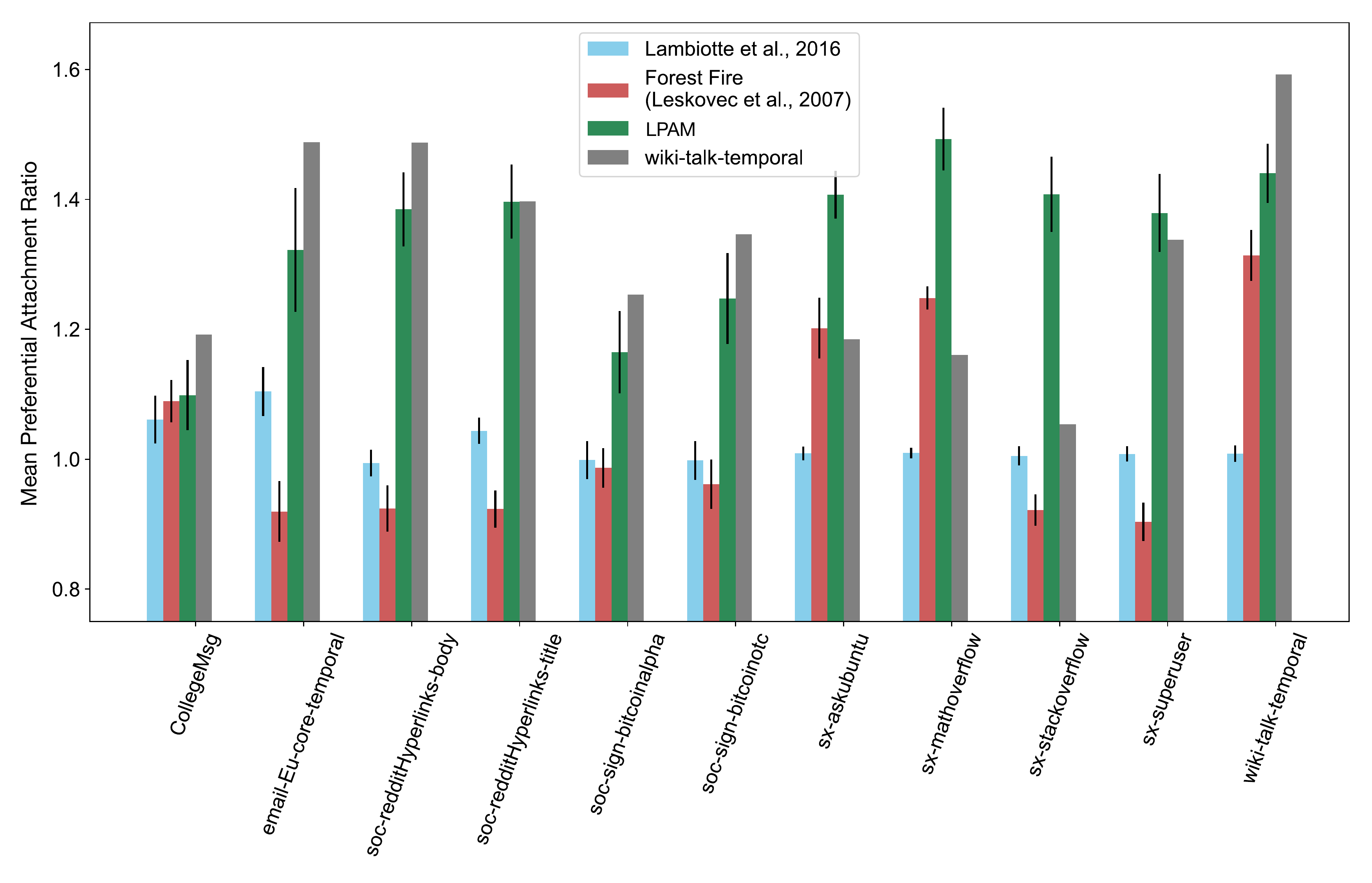}
    \centering
    \caption{Effect of preferential attachment. The y-axis shows the mean degree of a target's neighbors that a new node connects to, averaged across all nodes at a given timestep, divided by the mean degree of all of the target's neighbors, again averaged over all nodes at a given timestep. When this ratio is greater than 1, nodes preferentially connect to higher-degree neighbors. Empirical data (grey bars) are compared against the node copying mechanism (light blue) \cite{Bhat2016}, the Forest Fire model \cite{Leskovec2007}, and LPAM. Datasets are College messages (CollegeMsg) \cite{Panzarasa2009}, emails at a large European institution (email-Eu-core-temporal) \cite{paranjape_motifs_2017}, Reddit hyperlinks within the body of a Reddit post (soc-redditHyperlinks-body), or in the title (soc-redditHyperlinks-title) \cite{kumar2018community},  Bitcoin Alpha and Bitcoin OTC trust networks (soc-sign-bitcoinalpha and soc-sign-bitcoinotc) \cite{kumar2016edge,kumar2018rev2}, and conversations on Ask Ubuntu (sx-askubuntu), Math Overflow (sx-mathoverflow), Stack Overflow (sx-stackoverflow), and Stack Exchange Super User (sx-superuser) boards \cite{paranjape_motifs_2017}, and Wikipedia's talk pages (wiki-talk-temporal) \cite{paranjape_motifs_2017}. Error bars are standard errors of this ratio across all sampled network sizes. }
    \label{fig:pref_att}
\end{figure*}
\begin{figure}
    \includegraphics[width=\columnwidth]{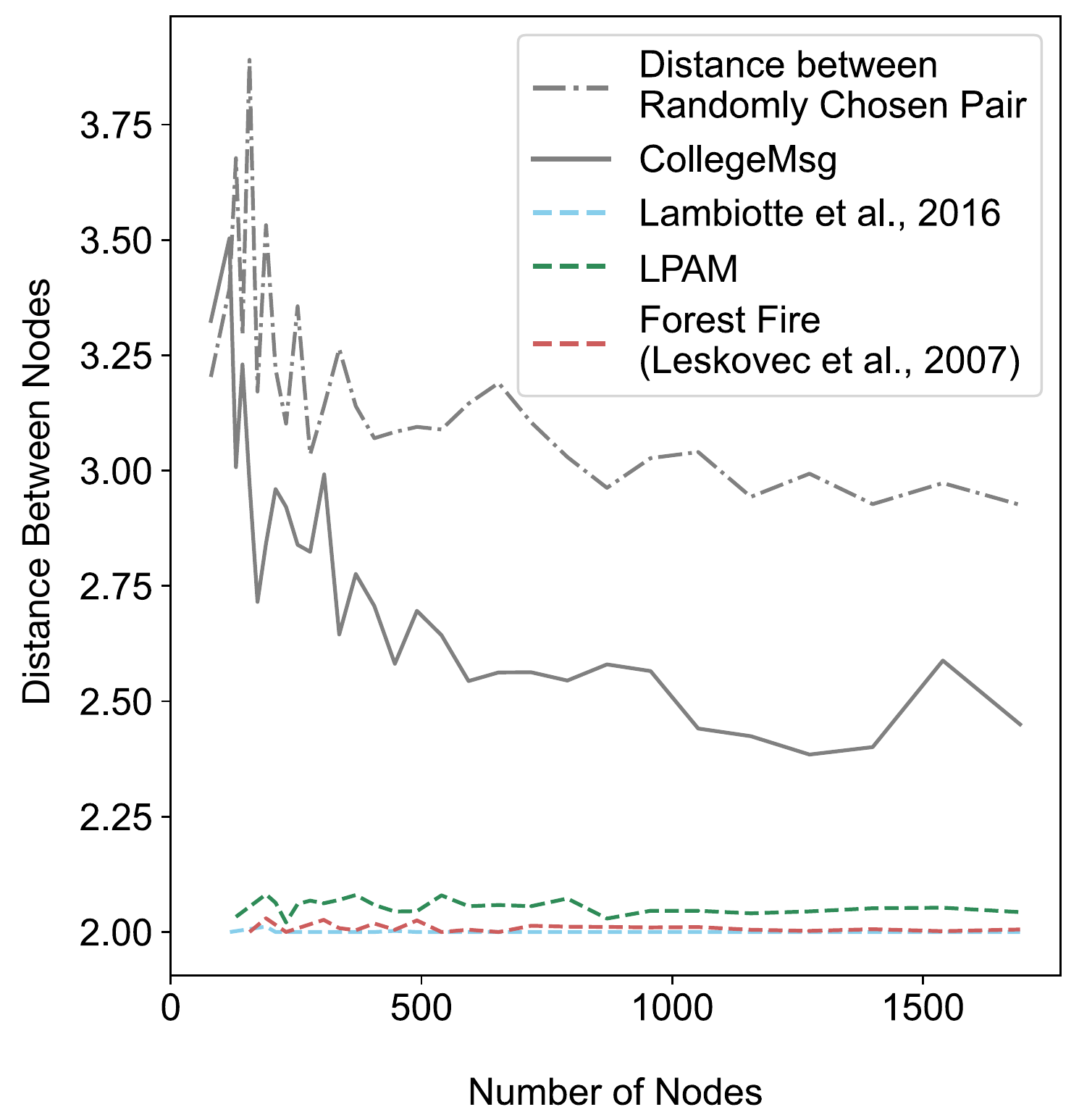}
    \centering
    \caption{Nodes make local connections. The distance between pairs nodes prior to forming a mutual connection with a new node. Distance between randomly chosen pairs in the College Messaging dataset \cite{Panzarasa2009}, and the empirical distance between nodes prior to connecting to a mutual new node. Other examples show similar results, see Supplementary Figure~{S5}.
    }
    \label{fig:distance}
\end{figure}
\subsection{Fitting Models}

Another methodological contribution of our work is fitting a clique densification model to empirical data of clique scaling. We measure the distribution of clique sizes for a given network size and compare this distribution to our model's prediction (see an example of this distribution in Supplementary Figure~S2). We find the parameters that fit the empirical distributions best across several network sizes, which can be characterized by maximizing the likelihood function averaged over the network sizes, $N$. We call this metric MeanMLE. Each $N$ are log-spaced steps between which the network grows 10\% until we reach the maximum network size. 
MeanMLE allows us to find parameters and models with the best overall fit to data, rather than the best at an arbitrary time point.

When fitting data, we discard model instances that will yield low likelihoods and remove models that time out computationally (take more than a few hours to run). We show in Supplementary Figure~S8 that each realization can have clique frequencies vary wildly for LPAM, and the wide variance can in turn can sometimes make calculating cliques computationally infeasible. This occurs rarely, however. For example, out of 150,000 instances across the three models used to fit Math Overflow, only 135 instances (0.09\%) are discarded. 

The LPAM, Forest Fire \cite{Leskovec2007}, and copying (Lambiotte et al., \cite{Lambiotte2016}) models all have parameters constrained to lie between 0 and 1. The entire parameter range is taken when models are fit and the parameters are randomly realized and rounded to the nearest 0.01, with 5 realizations on average for each parameter value. For the Forest Fire and LPAM, there are two parameters whose range is between 0 and 1, therefore there are $5\times 101 \times 101$ or approximately 50,000 realizations for each dataset. In contrast, for the copying model, there is only 1 parameter and therefore $5\times 101=505$  realizations.

\section{Results}\label{Results}

We compare the statistics of several empirical graphs against all candidate models: the Forest Fire model \cite{Leskovec2007}, which was the first of two models to explain densification, the copying model \cite{Lambiotte2016}, which provides theoretical predictions for clique scaling, and LPAM. While there are many other potential models one could compare against \cite{Dudin2014,zalanyi_properties_2003,paranjape_motifs_2017,leskovec_microscopic_2008}, our results show that LPAM captures basic aspects of network growth with a simple theoretically-grounded mechanism.

To test the importance of preferential attachment \cite{BA}, we measure the mean degree of the target node's neighbors to which a new node connects, divided by the mean degree of all the target node's neighbors, averaged over all network sizes sampled. Preferential attachment would imply that this ratio is greater than one. In Fig.~{\ref{fig:pref_att}}, we show our findings for all networks studied. While the copying model has a ratio of nearly one, implying no significant preferential attachment, the empirical data show a ratio significantly greater than one (strong preferential attachment) which is better captured with LPAM.  See Supplementary Figure~{S4} for further support of the consistency of these results across different empirical datasets.

We also plot the mean distance between nodes before they connect to each other, and compare this distance to a null model (connecting between random nodes) as well as to the candidate models, shown in Fig.~{\ref{fig:distance}} (similar plots are seen for other datasets in Supplementary Figure~{S5}). We find that nodes form links to nearby nodes (the distance is smaller than the null model), while the models assume even closer distances - neighbors of neighbors, implying a distance of 2. We therefore qualitatively capture the closeness of link formation, although the models tested do not fully address the links that are formed at a distance greater than 2. Capturing these nuances are left for future work.

Furthermore, we explore how the different mechanisms capture the scaling exponents of different clique sizes. We show in Fig.~{\ref{fig:fig1}}c that exponents increase significantly with clique size, which is qualitatively captured from the copying mechanism \cite{Bhat2016,Lambiotte2016} but this model has a lower exponent than what we find empirically. (This is also consistent with what we find in other datasets, shown in Supplementary Figure~{S6}.) 
LPAM, however, can better capture the scaling law exponents, and therefore help us understand why extremely dense cliques are unusually common in large networks. While the performance is comparable with the forest fire model, LPAM provides a clearer mechanism to explain this behavior. We also show in Supplementary Figure~S3 that LPAM captures the mean clique size better than the competing models. 

In order to determine the best overall model among these three, we take the mean {Kullback}-{Leibler} (KL) divergence \cite{Kullback1951} between model and empirical clique size distributions (details in Supplementary Figure~{S7}). We find that LPAM and the Forest Fire model have less error (lower KL divergence) than the copying model of Lambiotte et al. \cite{Lambiotte2016}, which suggests that the Lambiotte et al. model may not fully capture how networks grow. Although LPAM can sometimes outperform the other models, we do not claim that another model cannot fit data even better. The main goal of our paper is to instead provide a theoretically-motivated mechanism beyond the copying model. 

Finally, we can study ablation of LPAM either by removing node-copying or removing preferential attachment. Setting $r=0$, we remove traditional preferential attachment, and the model simplifies to the node copying model of Lambiotte et al. \cite{Lambiotte2016}, a poorer-fitting model. Alternatively, we can remove node copying and have nodes connect to other nodes preferentially based on degree. This simplifies LPAM to the Barabasi-Albert model \cite{BA}, whose degree is fixed independent of network size. Because neither simplification fits data as well, LPAM is an effective mechanism to reproduce the results we observe.


\section{Conclusion}
We observe that cliques scale super-linearly with network size, therefore we observe strong patterns in the higher-order structure of networks. Moreover, we observe that scaling exponents vary significantly for large and small cliques in a growing network. We further observe nodes connect locally (e.g., to neighbors of neighbors) and confirm previous analysis that nodes have preferential attachment. We develop a new mechanism, LPAM, to explain these patterns. LPAM is an extension of previous mechanisms in which a new node attaches to a target node and preferentially to the target node's higher-degree neighbors. We carried out an ablation study to show this is one of the simplest mechanisms to explain the empirical patterns we measure. 

There are a number of ways this method could be improved in future work. First, the mechanism is not theoretically grounded for cliques of order $k>2$. Next, LPAM does not fully match empirical data, which is both a disadvantage and an advantage in that it greatly simplifies the rich complex patterns that each observational network encodes. We notice in Supplementary Figure S7, for example, that LPAM performs worse than or similarly to the competing Forest Fire model for small networks, such as the College Messages or cryptocurrency networks. This points to finite size effects that our model overlooks. Even when the model performs well, LPAM's exponents are often lower than the empirical data (Supplementary Figure S6), and the simulated nodes connect to closer neighbors than in empirical data. This motivates extensions of LPAM to address finite size effects and the strong relation between clique size and scaling exponent. One way to improve this model, which might address some of its limitations, includes having new edges connect between two old nodes in the network with some probability, which is similar to the Newman-Watts small world model \cite{NEWMAN1999341}. Another way to improve the model could be to seed the model with a real network as an initial condition.
Finally, we assume that the fitted scaling laws are asymptotic, but this needs to be tested with more networks, especially with sizes in the hundreds of millions to billions (which our current computing power cannot tolerate).

\section{Code Availability}
The code is available at the following URL: \url{https://github.com/haochenpi314/Clique-Densification}.

\section{Supplementary Materials}
Below we show the robustness of our results across a range of datasets.
\setcounter{figure}{0}
\renewcommand\thefigure{S\arabic{figure}}    
\begin{figure*}
    \includegraphics[width=0.95\linewidth]{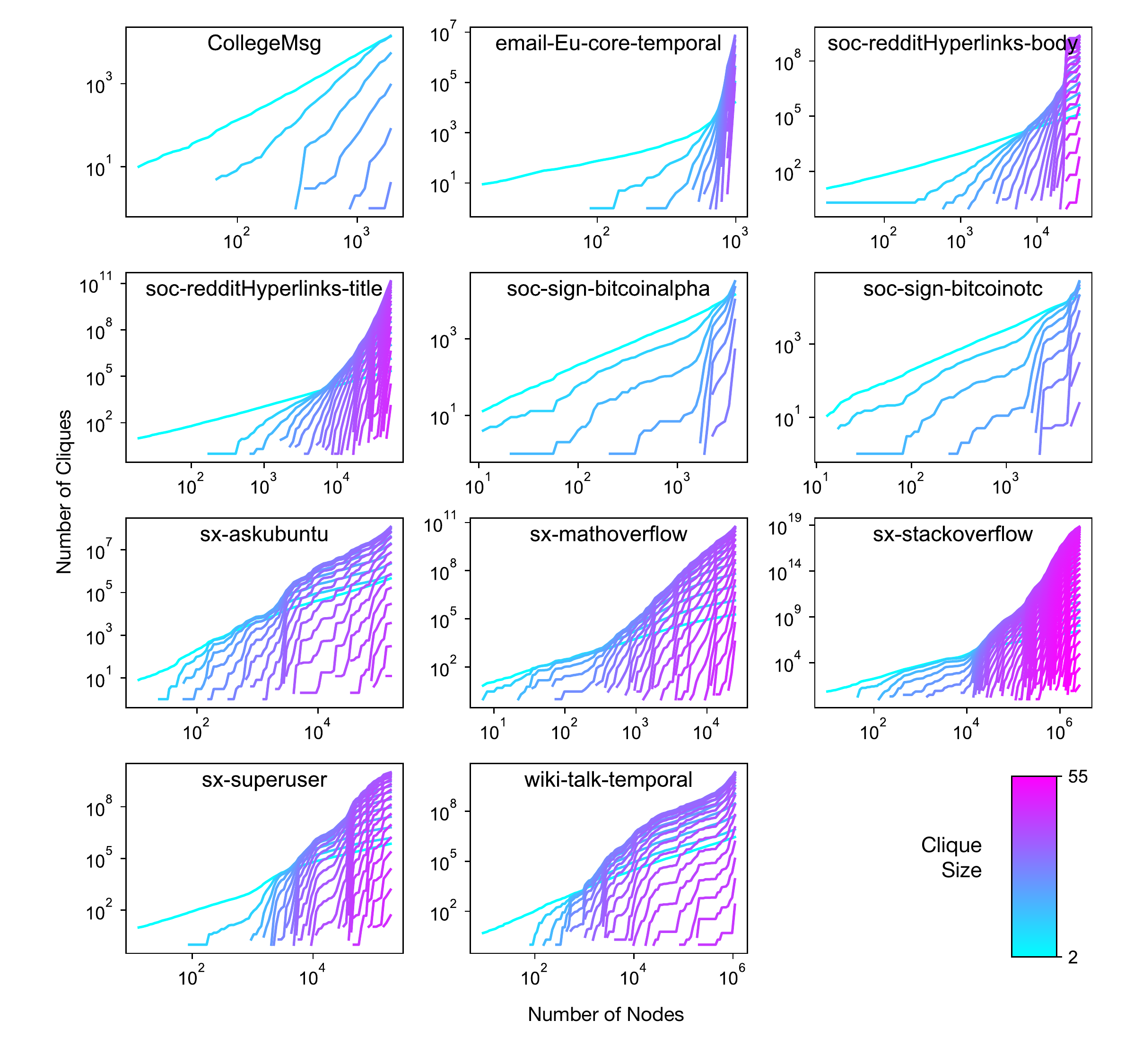}
    \centering
    \caption{The number of cliques versus network size in empirical data. Datasets are College messages (CollegeMsg) \cite{Panzarasa2009}, emails at a large European institution (email-Eu-core-temporal) \cite{paranjape_motifs_2017}, Reddit hyperlinks within the body of a Reddit post (soc-redditHyperlinks-body), or in the title (soc-redditHyperlinks-title) \cite{kumar2018community},  Bitcoin Alpha and Bitcoin OTC trust networks (soc-sign-bitcoinalpha and soc-sign-bitcoinotc) \cite{kumar2016edge,kumar2018rev2}, and conversations on Ask Ubuntu (sx-askubuntu), Math Overflow (sx-mathoverflow), Stack Overflow (sx-stackoverflow), and Stack Exchange Super User (sx-superuser) boards \cite{paranjape_motifs_2017}, and Wikipedia's talk pages (wiki-talk-temporal) \cite{paranjape_motifs_2017}.}
    \label{fig:allnumcliques}
\end{figure*}


\begin{figure*}[t]
    \includegraphics[width=\columnwidth]{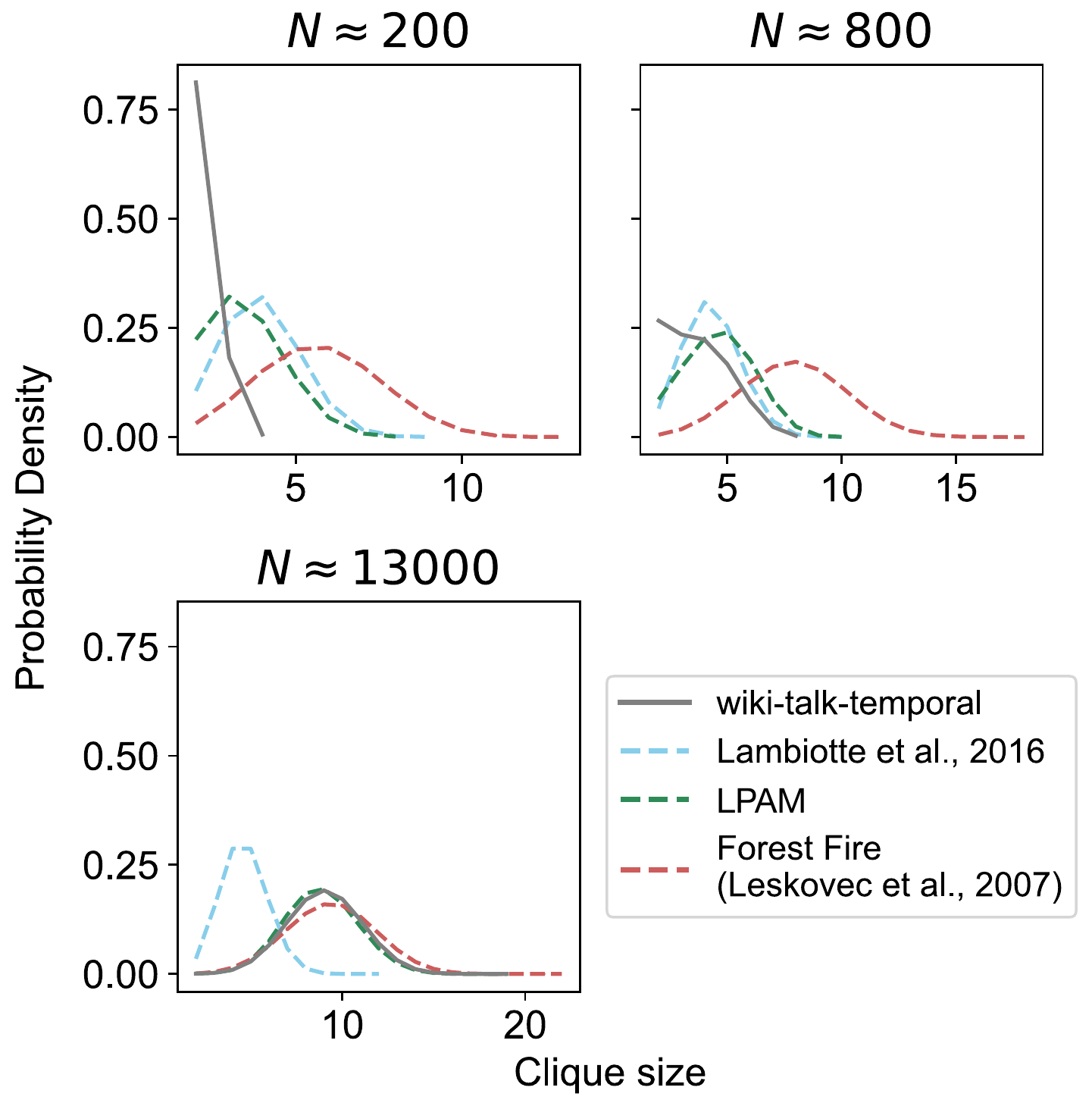}
    \centering
    \caption{A qualitative comparison of clique size distributions for varying network sizes. The empirical distributions from Wikipedia's talk pages (gray lines) \cite{paranjape_motifs_2017} moves to the right over time, and is better captured by LPAM and the Forest Fire models \cite{Leskovec2007} than Lambiotte et al. \cite{Lambiotte2016}.}
    \label{fig:dist_evo}
\end{figure*}

\begin{figure*}
    \includegraphics[width=2\columnwidth]{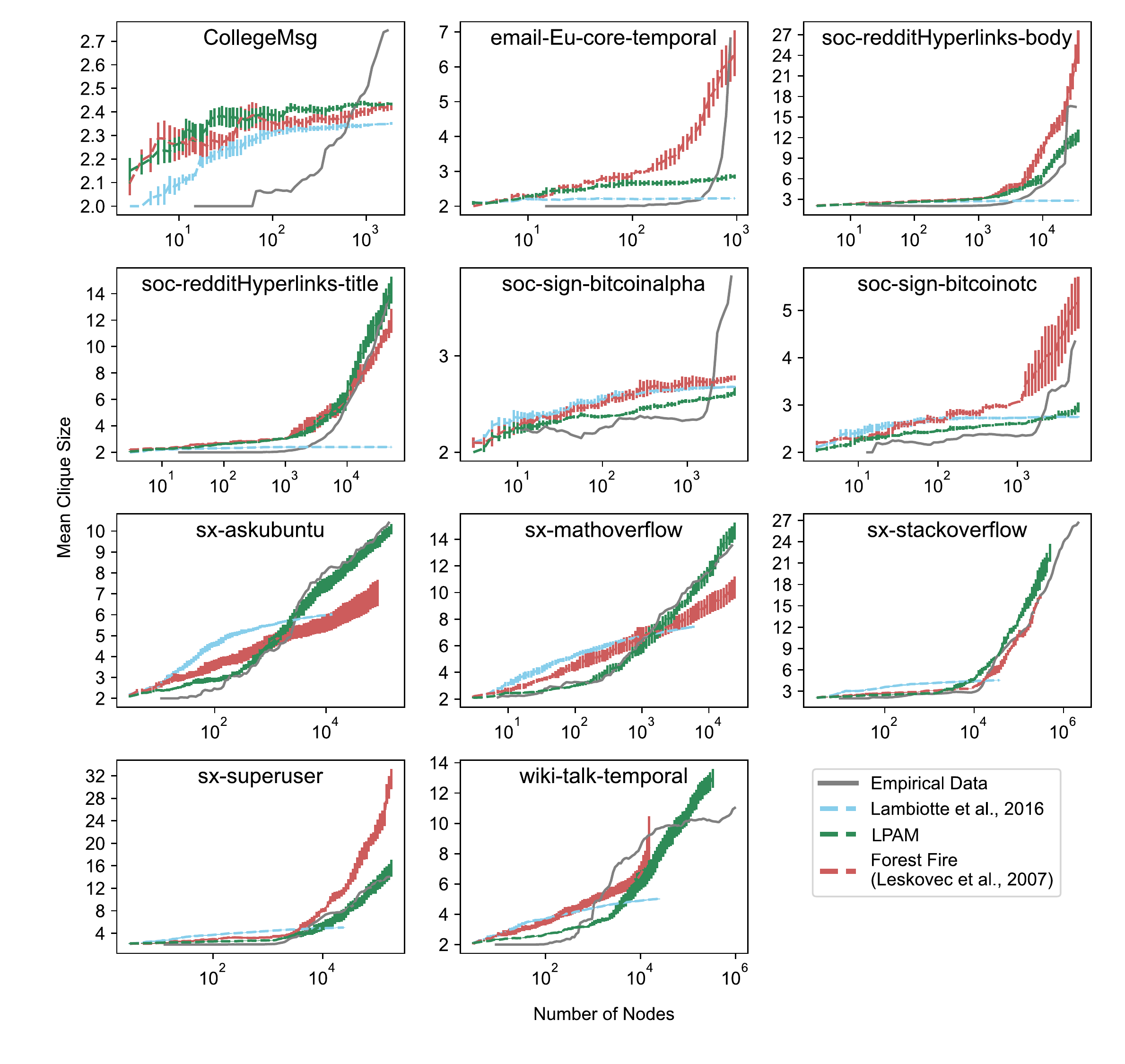}
    \centering
    \caption{
        Mean clique size versus network size for empirical data and models. Means and standard errors are from five model realizations on average. Datasets are College messages (CollegeMsg) \cite{Panzarasa2009}, emails at a large European institution (email-Eu-core-temporal) \cite{paranjape_motifs_2017}, Reddit hyperlinks within the body of a Reddit post (soc-redditHyperlinks-body), or in the title (soc-redditHyperlinks-title) \cite{kumar2018community},  Bitcoin Alpha and Bitcoin OTC trust networks (soc-sign-bitcoinalpha and soc-sign-bitcoinotc) \cite{kumar2016edge,kumar2018rev2}, and conversations on Ask Ubuntu (sx-askubuntu), Math Overflow (sx-mathoverflow), Stack Overflow (sx-stackoverflow), and Stack Exchange Super User (sx-superuser) boards \cite{paranjape_motifs_2017}, and Wikipedia's talk pages (wiki-talk-temporal) \cite{paranjape_motifs_2017}.}
    \label{fig:mean_clique_size_all}
\end{figure*}

\begin{figure*}
    \includegraphics[width=2\columnwidth]{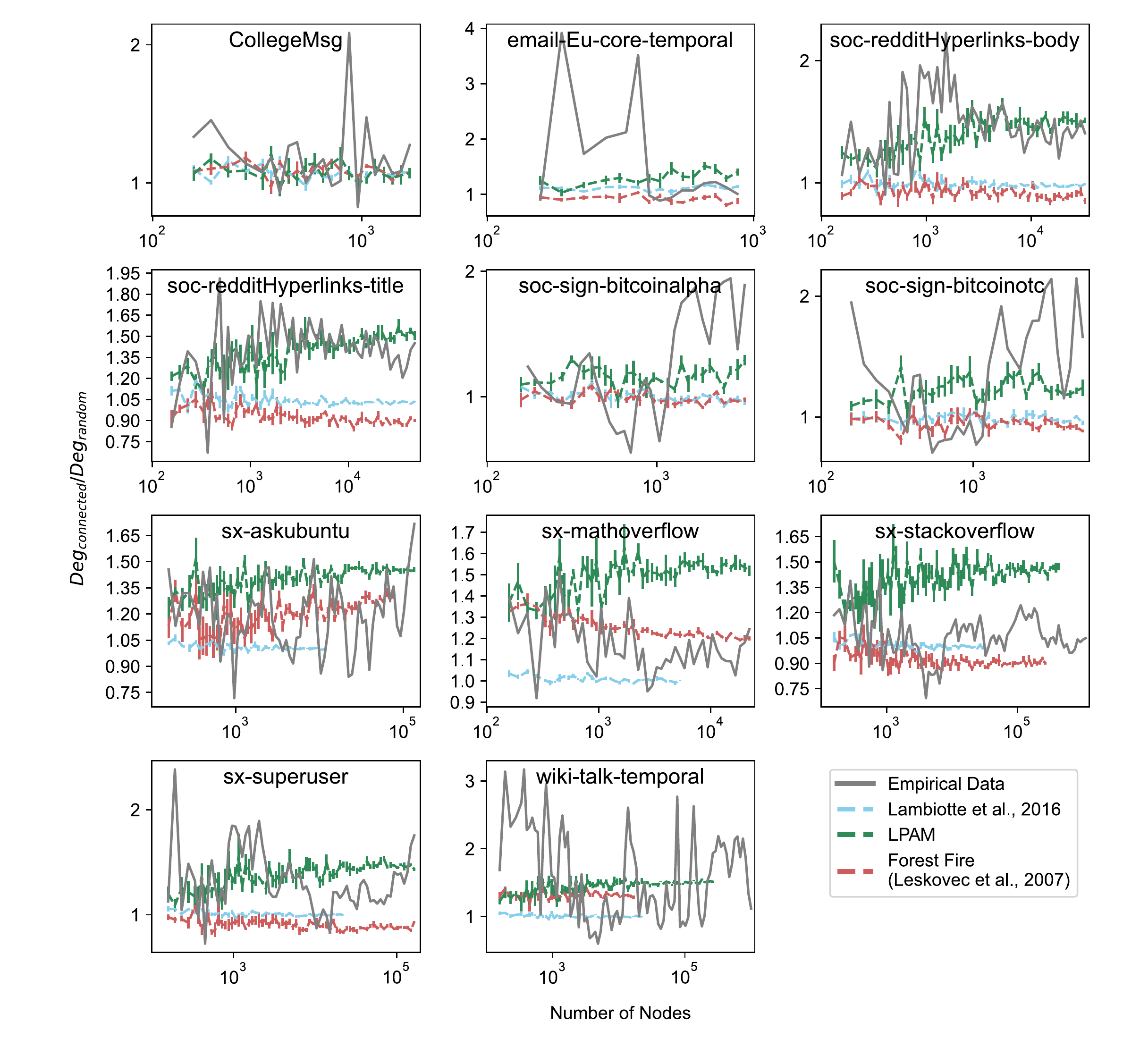}
    \centering
    \caption{Preferential attachment versus network size. The y-axis of each plot is the mean degree of newly connected nodes divided by the mean degree of random neighbors. Values greater than one indicate connected nodes have a higher degree than random neighbors, showing a preferential attachment (see Methods). In gray are empirical data, while blue, green, and red lines indicate the node copying model of Lambiotte et al. \cite{Lambiotte2016}, LPAM (our model), and the Forest Fire model \cite{Leskovec2007}, respectively. Datasets are College messages (CollegeMsg) \cite{Panzarasa2009}, emails at a large European institution (email-Eu-core-temporal) \cite{paranjape_motifs_2017}, Reddit hyperlinks within the body of a Reddit post (soc-redditHyperlinks-body), or in the title (soc-redditHyperlinks-title) \cite{kumar2018community},  Bitcoin Alpha and Bitcoin OTC trust networks (soc-sign-bitcoinalpha and soc-sign-bitcoinotc) \cite{kumar2016edge,kumar2018rev2}, and conversations on Ask Ubuntu (sx-askubuntu), Math Overflow (sx-mathoverflow), Stack Overflow (sx-stackoverflow), and Stack Exchange Super User (sx-superuser) boards \cite{paranjape_motifs_2017}, and Wikipedia's talk pages (wiki-talk-temporal) \cite{paranjape_motifs_2017}.
    }
    \label{fig:nbr_deg}
\end{figure*}

\begin{figure*}
    \includegraphics[width=2\columnwidth]{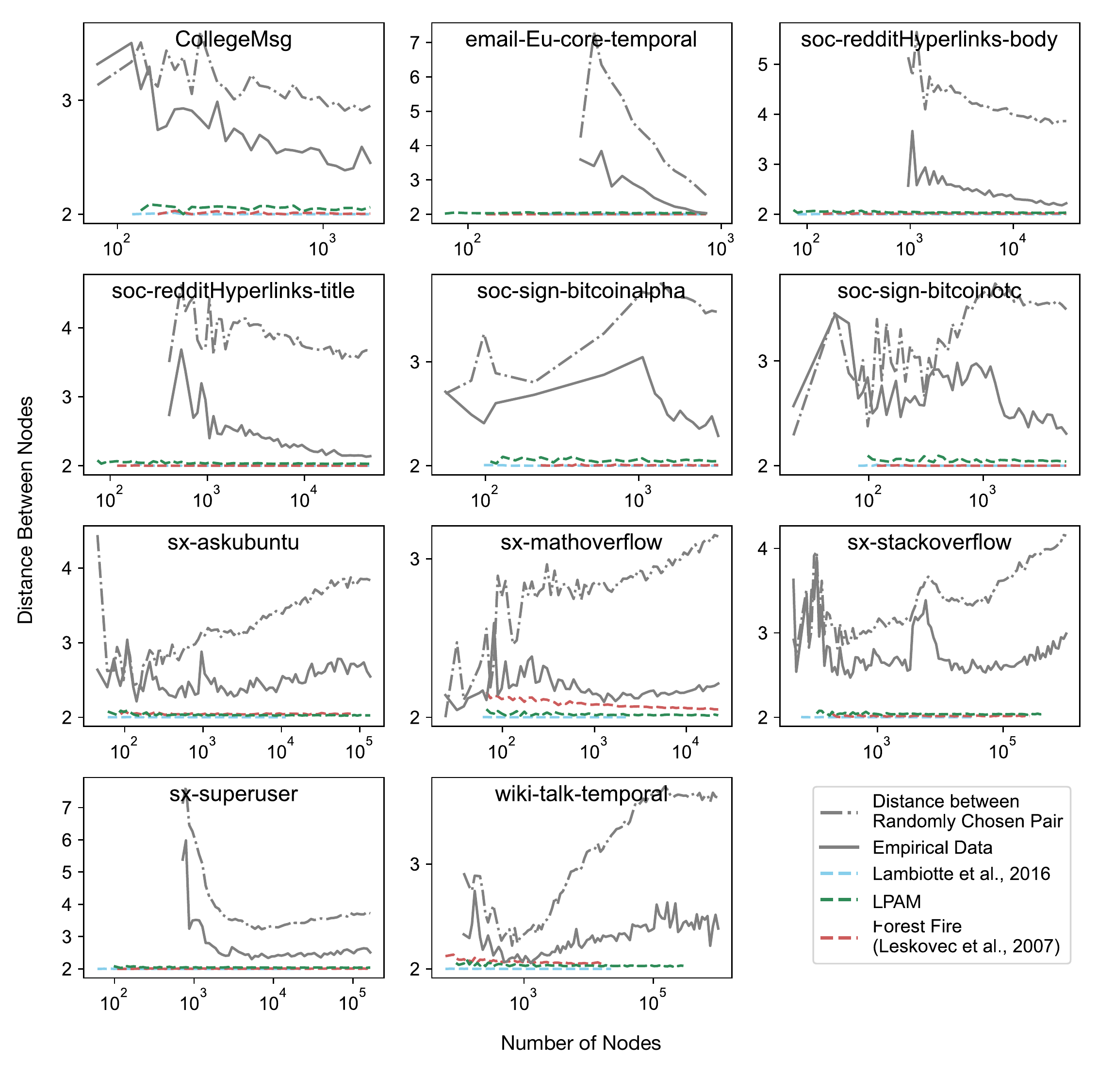}
    \centering
    \caption{Nodes form links locally. The geometric mean distance between newly formed links (solid gray line) is smaller than the geometric distance between links formed between random nodes (dashed gray line). Also shown are the node copying model of Lambiotte et al. \cite{Lambiotte2016}, LPAM (our model), and the Forest Fire model \cite{Leskovec2007}. Datasets are College messages (CollegeMsg) \cite{Panzarasa2009}, emails at a large European institution (email-Eu-core-temporal) \cite{paranjape_motifs_2017}, Reddit hyperlinks within the body of a Reddit post (soc-redditHyperlinks-body), or in the title (soc-redditHyperlinks-title) \cite{kumar2018community},  Bitcoin Alpha and Bitcoin OTC trust networks (soc-sign-bitcoinalpha and soc-sign-bitcoinotc) \cite{kumar2016edge,kumar2018rev2}, and conversations on Ask Ubuntu (sx-askubuntu), Math Overflow (sx-mathoverflow), Stack Overflow (sx-stackoverflow), and Stack Exchange Super User (sx-superuser) boards \cite{paranjape_motifs_2017}, and Wikipedia's talk pages (wiki-talk-temporal) \cite{paranjape_motifs_2017}.
    }
    \label{fig:alldistance}
\end{figure*}


\begin{figure*}
    \includegraphics[width=2\columnwidth]{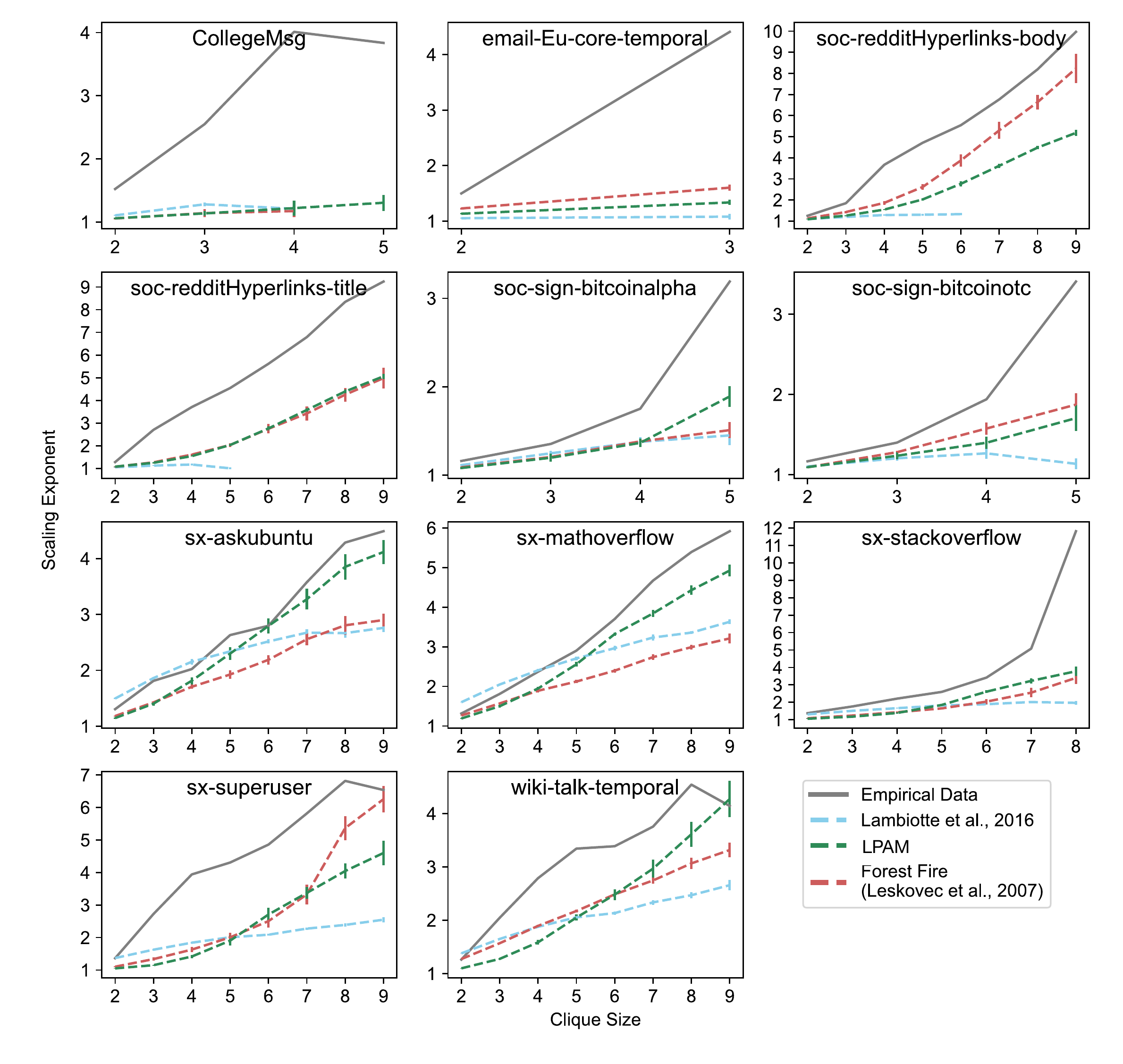}
    \centering
    \caption{Scaling exponents versus clique size. Scaling exponents are slope of the best fit line in the scaling plot, Fig.~\ref{fig:allnumcliques}, for each clique size. Datasets are College messages (CollegeMsg) \cite{Panzarasa2009}, emails at a large European institution (email-Eu-core-temporal) \cite{paranjape_motifs_2017}, Reddit hyperlinks within the body of a Reddit post (soc-redditHyperlinks-body), or in the title (soc-redditHyperlinks-title) \cite{kumar2018community},  Bitcoin Alpha and Bitcoin OTC trust networks (soc-sign-bitcoinalpha and soc-sign-bitcoinotc) \cite{kumar2016edge,kumar2018rev2}, and conversations on Ask Ubuntu (sx-askubuntu), Math Overflow (sx-mathoverflow), Stack Overflow (sx-stackoverflow), and Stack Exchange Super User (sx-superuser) boards \cite{paranjape_motifs_2017}, and Wikipedia's talk pages (wiki-talk-temporal) \cite{paranjape_motifs_2017}. Bars are standard errors across model realizations.
}
    \label{fig:scaling_law_all}
\end{figure*}



\begin{figure*}[t]
    \includegraphics[width=2\columnwidth]{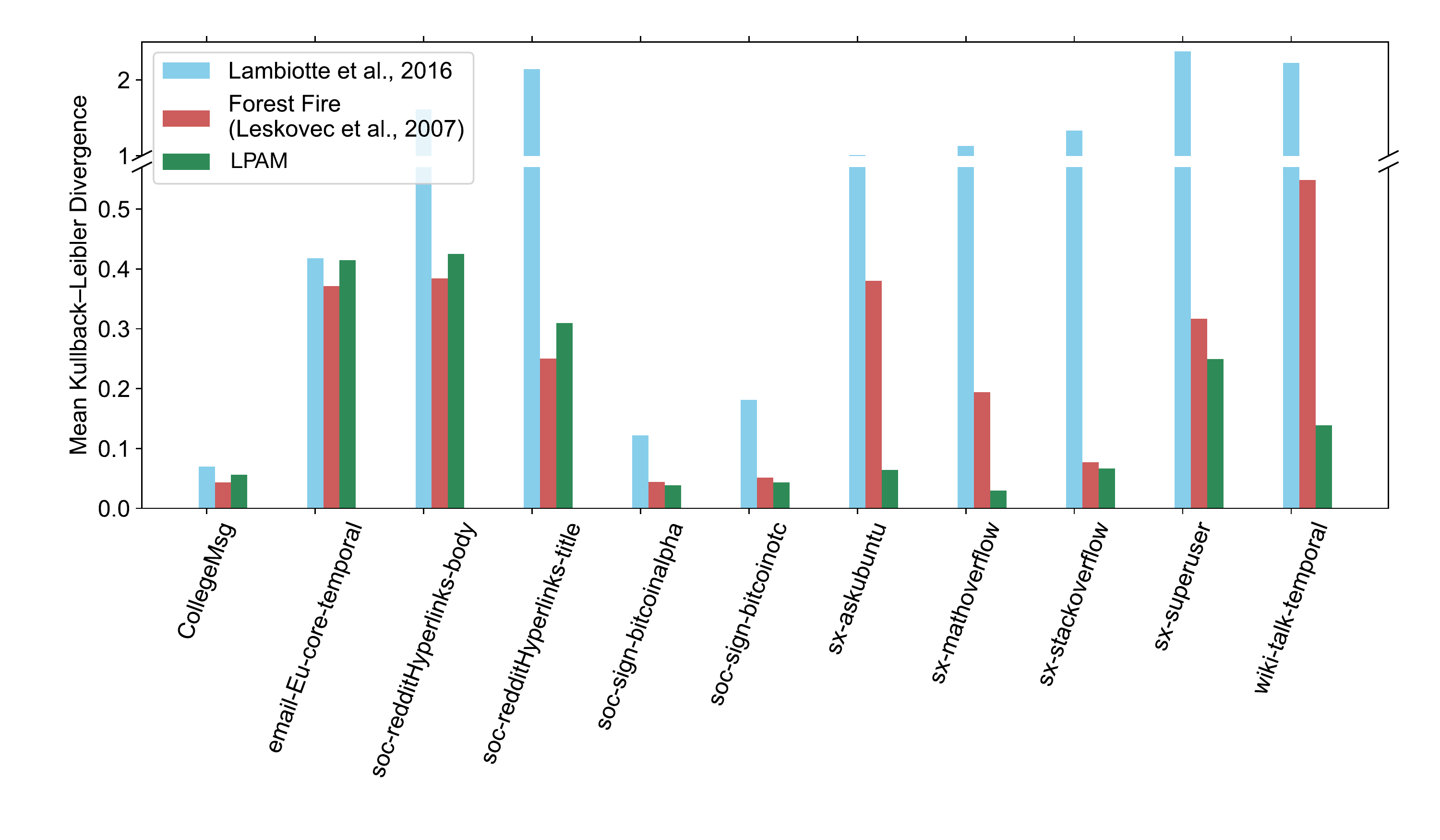}
    \centering
    \caption{Mean KL Divergence for each model across each dataset studied. Lower values indicate a better fit to data. Datasets are College messages (CollegeMsg) \cite{Panzarasa2009}, emails at a large European institution (email-Eu-core-temporal) \cite{paranjape_motifs_2017}, Reddit hyperlinks within the body of a Reddit post (soc-redditHyperlinks-body), or in the title (soc-redditHyperlinks-title) \cite{kumar2018community},  Bitcoin Alpha and Bitcoin OTC trust networks (soc-sign-bitcoinalpha and soc-sign-bitcoinotc) \cite{kumar2016edge,kumar2018rev2}, and conversations on Ask Ubuntu (sx-askubuntu), Math Overflow (sx-mathoverflow), Stack Overflow (sx-stackoverflow), and Stack Exchange Super User (sx-superuser) boards \cite{paranjape_motifs_2017}, and Wikipedia's talk pages (wiki-talk-temporal) \cite{paranjape_motifs_2017}.}
    \label{fig:bestfit}
\end{figure*}

\begin{figure*}[t]
    \includegraphics[width=2\columnwidth]{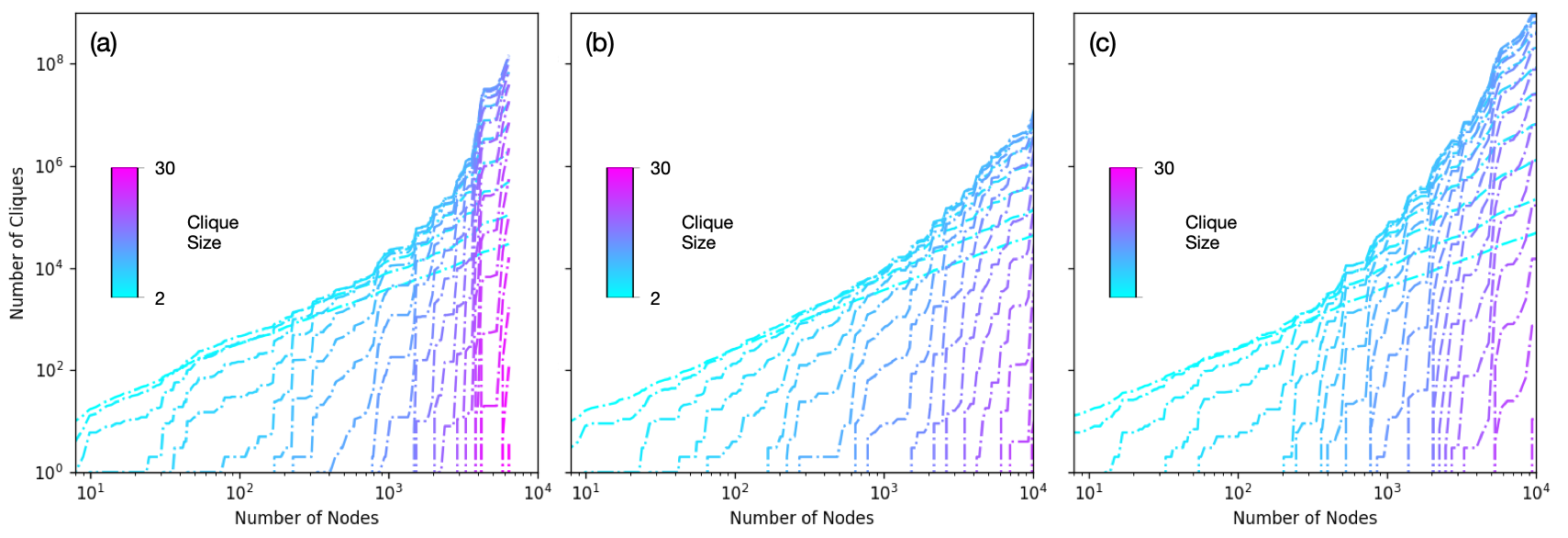}
    \centering
    \caption{Three realizations of the LPAM model with $p=0.42$ and $r=0.89$, which are the best-fit parameters for Math Overflow (sx-mathoverflow) \cite{paranjape_motifs_2017}. We notice a consistent scaling law, as demonstrated in the low standard deviation of exponents in main text Fig. 1c, but the absolute number of cliques can vary significantly for each realization.}
    \label{fig:sim}
\end{figure*}

\clearpage
%

\end{document}